%Paper: hep-th/9310140
%From: fteorica@cpd.uva.es
%Date: Thu, 21 OCT 93 13:45 GMT

%%%%%%%%%%%%%%%%%%%%% AMS-TEX FILE %%%%%%%%%%%%%%%%%%%%%

\magnification\magstep1
\TagsOnRight
\baselineskip=12pt
\NoBlackBoxes
\font\ninepoint=cmr9
\font\titulo=cmbx10 scaled\magstep2

\def\Dr{1}
\def\Ji{2}
\def\Li{3}
\def\Lii{4}
\def\Liii{5}
\def\Liv{6}
\def\Ti{7}
\def\BHOS{8}
\def\BHOSii{9}
\def\IW{10}
\def\BLL{11}
\def\Gill{12}
\def\TesHe{13}
\def\SHO{14}
\def\HOS{15}
\def\HMOS{16}
\def\Abe{17}
\def\Maj{18}

            % los nmeros reales
            % los nmeros enteros
\def\k{\kappa}                 % coeficientes
                  % coef puntos
                  % coef lineas
\define\g{{\frak g}}           % algebra de Lie de G
\define\J{{\Bbb J}}            %  gen. contrado
\define\K{{\Bbb K}}            %  gen. contrado
\define\He{{\Bbb H}}           %  gen. contrado
\define\Pe{{\Bbb P}}           %  gen. contrado
           % dualidad 2d
            %  extension
          %
\redefine\mdot{\!\cdot\!}
\redefine\C{\text{\ \!C}}
\redefine\Sk{\text{\ \!S}}
\define\>#1{{\bold#1}}                 %  notacin para vectores
\define\co{\Delta}                     %  coproducto
\define\conm#1#2{\left[ #1,#2 \right]} %  un conmutador
\def\1{\'{\i}}                         %  i con acento
\define\q#1{{\left[ #1\right]}_q}      %  un q-numero [x]q
\define\qq#1{{\left\{ #1\right\} }_q}  %  un q-numero {x}q

\def\picture #1 by #2 (#3){\vbox to #2{
       \hrule width #1 height 0pt depth 0pt
       \vfill \special{picture #3}}}

%%%%%%%%%%%%%%%%%%%%%%%%%%%%%%%

\
\vskip 3truecm

\centerline {\titulo  4D QUANTUM AFFINE ALGEBRAS}
\bigskip
\centerline {\titulo  AND SPACE--TIME q-SYMMETRIES}
\bigskip

\vskip 1truecm

\centerline {A. Ballesteros, F.J. Herranz, M.A. del Olmo and M. Santander}
\bigskip\bigskip
\centerline{\it Departamento de F\1sica Te\'orica, Universidad de
                Valladolid.}
\centerline{\it E--47011 Valladolid, Spain.}
\smallskip
\centerline{e-mail: fteorica\@cpd.uva.es}

\vskip 3truecm

{\ninepoint
\noindent {\bf Abstract.}
A global model of $q$-deformation for the quasi--orthogonal Lie algebras
generating the groups of motions of the four--dimensional affine
Cayley--Klein geometries is obtained starting from the three dimensional
deformations. It is shown how the main algebraic classical properties of the
CK systems can be implemented in the quantum case. Quantum deformed versions
of either the space--time or space symmetry algebras (Poincar\'e (3+1),
Galilei (3+1), 4D Euclidean as well as others) appear in this context as
particular cases and several $q$-deformations for them are directly
obtained.}

\vskip 1truecm \centerline {PACS: 02.10.T; 02.20.Sv; 03.65.Fd}
\bigskip
\medskip

\vskip 1truecm \centerline {October, 1993}

\vfill\eject

\centerline {\bf I. INTRODUCTION.}
\medskip
\medskip

The quantum analogues of  real non--simple algebras such as
Poincar\'e and Galilei ones have been obtained from the Drinfel'd--Jimbo
deformation of the classical series of Lie algebras [\Dr,\Ji] in a two-step
process [\Li-\Ti]: first, a quantum real form is  deduced from a
quantum complex one and, afterwards, a ``quantum contraction" is applied on
it. In two recent works [\BHOS,\BHOSii] a global scheme of quantum
deformations of the three and six dimensional Lie algebras generating the
groups of motions of the two and three dimensional classical Cayley--Klein
geometries (CKG) has been introduced. The interest of this approach to
quantization is based in the following facts:
\smallskip

\item {1)} The group of motions of the $N$-dimensional CKG include the
orthogonal/pseudoorthogo-nal groups $SO(p,q)$, $(p+q=N+1,\, p\ge q\ge 0)$
and their In\"on\"u--Wigner (IW) contractions [\IW], i.e., the so called
quasi--simple groups of orthogonal type. Relevant kinematical
groups [\BLL] appear among them; for instance, De Sitter,  Poincar\'e,
Galilei,  etc.

\smallskip

\item {2)} This scheme is self--contained in the sense that no
reference to complex forms is needed: it is worked out from the very
beginning with the CK Lie algebras that are real forms by themselves.

\smallskip

\item {3)} In spite of the fact that after ``quantization" the purely
geometrical meaning of the CK structure disappears, the main part of the
associated algebraic structure can be generalized and remains relevant. For
instance, the $q$-versions of involutions and contractions relating
the different quantum CK algebras are easily deduced.
\medskip

In this paper we present the first step towards the generalization of this
CK quantization to higher dimensions. A remarkable feature is that the
classical CKG  contain, in any dimension, lower dimensional ones as
essential constituents. We use this fact to propose, starting from the three
dimensional case  [\BHOSii], a quantum algebra associated to the four
dimensional affine CKG. An Ansatz for the Hopf structure is immediately
written in terms of the 3D case and the algebra relations are found by
requiring the coproduct to be a Hopf algebra homomorphism. The (3+1)
Poincar\'e and (3+1) Galilei algebras,  as well as the 4D Euclidean algebra,
are 4D affine CK algebras. We get for them some new $q$-deformations and we
also recover known results, as the $\k$-Poincar\'e of Lukierski {\sl et
al.} [\Liv]. It is interesting to note that the $\k$-Poincar\'e is directly
obtained in its final form [\Liv] as a particular case of the general CK
algebra, so no use of any non--linear change of basis [\Gill] is made.

\medskip

In Section II we give a brief outline of the
structure of the classical 4D affine CK algebras. The inductive method
giving rise to the (global) $q$-deformation is presented in Section III; it
uses known classical and quantum properties related to lower dimensional CK
algebras. Section IV is devoted to the discussion of the $q$-kinematical
algebras contained within the general quantum affine CK scheme.

\bigskip
\bigskip

\centerline {\bf II. CLASSICAL AFFINE 4D-CAYLEY--KLEIN ALGEBRAS.}
\medskip\medskip

The CK algebras associated to the 4D-CK geometries are 10-dimensional real
Lie algebras characterized by four fundamental real parameters
$(\k_1,\k_2,\k_3,\k_4)$ [\TesHe-\HOS]. These parameters describe the
kind of measure of separation between points $(\k_1)$, lines $(\k_2)$,
planes $(\k_3)$ and hyperplanes $(\k_4)$, so, if $\k_i$ is $>0$, $=0$ or
$<0$ the measure is elliptical, parabolical or hyperbolical, respectively.
When all the $\k_i$ are non zero we obtain the $so(p,q)$ $(p+q=5,\, p\ge
q\ge 0)$ algebras and for some $\k_i=0$ we get quasi--simple ones. In
particular, the case $\k_1=0$ (parabolic measure of distance between points
and zero constant curvature) gives rise to the {\sl {affine}} CK Lie
algebras which are globally denoted as $\g_{ (0,\k_2,\k_3,\k_4)}$. In the
basis  $\{ P_i,J_{ij}\}$ ($i,j=1,2,3,4;\ i<j$), the Lie brackets of

$\g_{(0,\k_2,\k_3,\k_4)}$ are given by $$
\aligned [P_i,P_j] &=0 ,\qquad
[J_{ij},P_k] =\delta_{ik}P_j-\delta_{jk}\k_{ij}\,P_i ,\\
[J_{ij},J_{lm}]&=\delta_{im}J_{lj}-\delta_{jl}J_{im}+
\delta_{jm}\k_{lm}J_{il}+\delta_{il}\k_{ij}J_{jm},\quad\ i\le l,\ j\le m;
\endaligned
\tag2.1
$$
where
$$
\k_{ij}=\k_{i+1}\k_{i+2}\dots\k_j,\quad i,j=1,2,3,4;\quad i<j.\tag2.2
$$

{}From relations (2.1) one can easily prove that the CK
groups  $G_{(0,\k_2,\k_3,\k_4)}$ generated by $\g_{(0,\k_2,\k_3,\k_4)}$
have a semidirect product structure: $ G_{(0,\k_2,\k_3,\k_4)}\equiv
T_4\odot G_{(\k_2,\k_3,\k_4)}, $ where $T_4=\langle P_i \rangle$ is an
abelian subgroup and $G_{(\k_2,\k_3,\k_4)}=\langle J_{ij} \rangle$ is a CK
subgroup.

\medskip

A differential realization of $\g_{(0,\k_2,\k_3,\k_4)}$ on $\Bbb L^2 (\Bbb
R^4)$ is given by
$$
P_i=\partial_i,\qquad  J_{ij}=\k_{ij}x_j\partial_i  -  x_i\partial_j.
\tag2.3
$$

The centre of $U\g_{(0,\k_2,\k_3,\k_4)}$ is generated by the
second and fourth order Casimir operators which read (see Appendix
A):
$$
\align
\Cal C_1&=\k_2\k_3\k_4 P_1^2+\k_3\k_4
P_2^2+\k_4P_3^2+P_4^2,\tag2.4\\
\Cal C_2&=\k_2 {W_1}^2+{W_2}^2+\k_3
{W_3}^2+\k_3\k_4 W_4^2,\tag2.5
\endalign
$$
with
$$
\alignedat2
W_1&=-(\k_3P_2J_{34}-P_3J_{24}+P_4J_{23}),&\quad
W_2&=\k_2\k_3P_1J_{34}-P_3J_{14}+P_4J_{13},\\
W_3&=-(\k_2P_1J_{24}-P_2J_{14}+P_4J_{12}),&\quad
W_4&=\k_2P_1J_{23}-P_2J_{13}+P_3 J_{12}.
\endalignedat
\tag2.6
$$

There exists a set of four basic commuting involutions $\Theta_{(i)}$
$(i=0,1,2,3)$ in $\g_{ (0,\k_2,\k_3,\k_4)}$. Each involution $\Theta_{(i)}$
determines the Lie subalgebra  $\frak h^{(i)}\subset \g$ of all elements of
$\g$ invariant under $\Theta_{(i)}$,
$$
\aligned
\frak h^{(0)}&=\langle J_{12},J_{13},J_{14},J_{23},J_{24},J_{34}\rangle,
\quad \frak h^{(1)}=\langle P_1,J_{23},J_{24},J_{34}\rangle,\\
\frak h^{(2)}&=\langle P_1,P_2,J_{12},J_{34}\rangle,\quad\
\frak h^{(3)}=\langle P_1,P_2,P_3,J_{12},J_{13},J_{23}\rangle .
\endaligned
\tag2.7
$$

 The dimensions of these subalgebras are 6, 4, 4 and
6 for $i=0,1,2,3$, respectively. Therefore, the homogeneous spaces
$\Cal X^{(i)}\equiv G/H^{(i)}$, where $H^{(i)}$ is the Lie subgroup of $G$
associated to $\frak h^{(i)}$, are symmetric and  $G$ acts transitively
on each of them. Moreover, they are identified with the spaces of points,
lines, planes and hyperplanes in the CKG according to $i=0,1,2,3$.

The action of the involution $\Theta_{(i)}$ over a generic element $X$ in
the basis $\langle P_i,J_{ij}\rangle$ of $\g$ is:
$$
\Theta_{(i)}(X)=\left\{
\aligned
&X\quad \text {if}\ X\in \frak h^{(i)}\\
-&X\quad \text {if}\ X\notin \frak h^{(i)}
\endaligned
\right. ,\quad i=0,1,2,3;
\tag2.8
$$
and extends by linearity to the whole algebra. These involutions generate
a $\Bbb Z_2^{\otimes 4}$  abelian group, giving rise to a grading of

$\g_{(0,\k_2,\k_3,\k_4)}$ \cite{\HMOS}.  \medskip

The 4D affine CKG are related by means of IW contractions. By applying the
transformation
$$
\Gamma_{(i)}(X)=\left\{
\aligned
&X\quad \text {if}\ X\in \frak h^{(i)}\\
\varepsilon&X\quad \text {if}\ X\notin \frak h^{(i)}
\endaligned
\right. ,\quad i=1,2,3;\tag2.9
$$
over the basis elements, computing the new Lie commutators   and
making the limit $\varepsilon\to 0$, we obtain a new CK algebra with the
parameter $\k_{i+1}=0$. Its associated 4D affine CKG describes the behaviour
of the original one in the neighbourhood of a  line, plane or hyperplane,
respectively.
\medskip

Within a given 4D affine CK algebra there are four 3D affine CK
subalgebras (linked to coordinate hyperplanes) that are described by means
of three real parameters depending on the fundamental $\k_i$. They are:
$$
\alignedat2
\frak h^{(3)}\equiv\Pi_{123}&  =\langle
P_1,P_2,P_3,J_{12},J_{13},J_{23}\rangle,
 &\qquad &(0,\k_2,\k_3);\\
\Pi_{124}&=\langle
P_1,P_2,P_4,J_{12},J_{14},J_{24}\rangle, &\qquad
&(0,\k_2,\k_3\k_4);\\
\Pi_{134}&=\langle
P_1,P_3,P_4,J_{13},J_{14},J_{34}\rangle, &\qquad
&(0,\k_2\k_3,\k_4);\\
\Pi_{234}&=\langle
P_2,P_3,P_4,J_{23},J_{24},J_{34}\rangle, &\qquad
&(0,\k_3,\k_4).
\endalignedat
\tag2.10
$$
These subalgebras will play a central role for getting the quantum
deformation. Note also that $\Pi_{123}$ ($\g_{(0,\k_2,\k_3)}$) contains
exactly the same generators as the affine case $N=3$ described in [\BHOSii].
\bigskip
\bigskip

\centerline {\bf III. QUANTUM AFFINE ALGEBRAS.}
\medskip
\medskip

The pattern of $q$-deformation ({\sl {\`a la}} Drinfel'd--Jimbo [1,2]) we
are proposing for the 4D affine CK algebras is inspired in the same leading
idea as the 2D and 3D cases were [\BHOS,\BHOSii]: to construct a global
model of quantization (simultaneous description of the Hopf algebra
structure [\Abe,\Maj]) for the 4D affine CK algebras such that for any
tetrad $(0,\k_2,\k_3,\k_4)$ there always exists some deformed
Lie brackets at the algebra level. This deformation can be deduced from the
quantum 3D affine case $U_q\g_{(0,\k_2,\k_3)}$ [\BHOSii] by using the
embedding (2.10) as follows.

\medskip

First, we write the coproducts for
$\Pi_{124}$, $\Pi_{134}$ and $\Pi_{234}$ by considering these subalgebras as
quantum 3D affine algebras (see Appendix B). Note that all of them are
expected to resemble explicitly the four dimensional structure since they
contain the new parameter $\k_4$ as well as the $P_4$ affine translation.
Afterwards, we assume that 1) these
coproducts are restrictions  of the general coproduct for $U_q\frak
g_{(0,\k_2,\k_3,\k_4)}$ and 2) no more terms than the ones arising in these
restrictions appear in the four dimensional comultiplication.
Restriction means that the generators of $U_q\frak
g_{(0,\k_2,\k_3,\k_4)}$  not included in a given
subalgabra are necessarily taken as zero.

\medskip

In order to illustrate our procedure we consider, for instance, the
generator $J_{14}$ which belongs to $\Pi_{124}$ and $\Pi_{134}$.
{}From (B.3) and (B.5) of Appendix B we write:
$$
\align
\co(J_{14})\big|_{\Pi_{124}}&=e^{-{z\over 2}P_4}\otimes
J_{14} + J_{14} \otimes e^{{z\over2}P_4} + \tfrac z 2 J_{12}e^{-{z\over
2}P_4} \otimes \k_3\k_4 P_2 - \k_3\k_4 P_2
\otimes e^{{z\over2}P_4}\tfrac z 2 J_{12},\\
\co(J_{14})\big|_{\Pi_{134}}&=e^{-{z\over 2}P_4}\otimes
J_{14} + J_{14} \otimes e^{{z\over2}P_4} + \tfrac z 2 J_{13}e^{-{z\over
2}P_4} \otimes \k_4 P_3 - \k_4 P_3
\otimes e^{{z\over2}P_4}\tfrac z 2 J_{13}.
\endalign
$$
Following our Ansatz, the coproduct for $J_{14}$ would be
$$
\align
\co(J_{14})&=e^{-\tfrac z2 P_4}\otimes J_{14} +J_{14}\otimes e^{\tfrac z2
P_4} +\tfrac z2 J_{12}e^{-\tfrac z2 P_4}\otimes \k_3\k_4 P_2
- \k_3\k_4   P_2 \otimes   e^{\tfrac z2 P_4} \tfrac z2 J_{12}\\
&\qquad\qquad\qquad\qquad
+\tfrac z2 J_{13}e^{-\tfrac z2 P_4}\otimes \k_4 P_3
- \k_4   P_3 \otimes   e^{\tfrac z2 P_4} \tfrac z2 J_{13}.
\endalign
$$

In this way, we obtain for all the generators the coproduct proposed in
(3.1). It is easy to check the coassociativity; counit and
antipode are  derived from (3.1). On the other hand, the
commutation relations compatible with this coproduct has still to be
determined. To do this we follow the same procedure, i.e., we start with the
restrictions of the deformed Lie brackets to $\Pi$ subalgebras, which are
explicitly given in Appendix B. It is remarkable that the mere addition of
an extra term in three of the deformed Lie brackets implies that the
required homomorphism condition is fulfilled (compare (B.4,6,8) with
(3.4c)). The classical limit ($z\to 0$) of the whole quantum structure can
be also straightforwardly checked.

\medskip

We conclude that the quantized universal enveloping affine CK algebra
$U_q\frak g_{(0,\k_2,\k_3,\k_4)}$ is  defined by:
\medskip

\noindent
{\bf a)} {\sl\underbar {The Hopf structure}}.
\medskip

\noindent
{i) Coproduct:}
$$
\aligned
&\qquad\qquad\qquad\co(X)=1\otimes X+ X\otimes 1,\quad
X\in\{P_4;J_{12},J_{13},J_{23}\},\\ &\qquad\qquad\qquad\co(P_i)=e^{-\frac z2
P_4}\otimes P_i +P_i\otimes e^{\frac z2 P_4},  \quad i=1,2,3.\\
\co(J_{14})&=e^{-\tfrac z2 P_4}\otimes J_{14} +J_{14}\otimes e^{\tfrac z2
P_4} +\tfrac z2 J_{12}e^{-\tfrac z2 P_4}\otimes \k_3\k_4 P_2
- \k_3\k_4   P_2 \otimes   e^{\tfrac z2 P_4} \tfrac z2 J_{12}\\
&\qquad\qquad\qquad\qquad
+\tfrac z2 J_{13}e^{-\tfrac z2 P_4}\otimes \k_4 P_3
- \k_4   P_3 \otimes   e^{\tfrac z2 P_4} \tfrac z2 J_{13},\\
\co(J_{24})&=e^{-\tfrac z2 P_4}\otimes J_{24} +J_{24}\otimes e^{\tfrac z2
P_4} -\tfrac z2 J_{12}e^{-\tfrac z2 P_4}\otimes \k_3\k_4 P_1
+\k_3\k_4  P_1 \otimes   e^{\tfrac z2 P_4}\tfrac z2  J_{12}\\
&\qquad\qquad\qquad\qquad
+\tfrac z2 J_{23}e^{-\tfrac z2 P_4}\otimes \k_4 P_3
- \k_4 P_3 \otimes   e^{\tfrac z2 P_4} \tfrac z2   J_{23},\\
 \co(J_{34})&=e^{-\tfrac z2 P_4}\otimes J_{34} +J_{34}\otimes e^{\tfrac z2
P_4} -\tfrac z2 J_{13}e^{-\tfrac z2 P_4}\otimes \k_4 P_1
+\k_4  P_1 \otimes   e^{\tfrac z2 P_4}\tfrac z2  J_{13}\\
&\qquad\qquad\qquad\qquad
-\tfrac z2 J_{23}e^{-\tfrac z2 P_4}\otimes\k_4 P_2
+ \k_4  P_2 \otimes   e^{\tfrac z2 P_4} \tfrac z2 J_{23}.
\endaligned\tag3.1
$$

\noindent
{ii) Counit:}
$$
\epsilon(P_i)=\epsilon(J_{ij})=0,\quad i,j=1,2,3,4.\tag3.2
$$

\noindent
{iii) Antipode:}
$$
\gamma(X)=-e^{\tfrac {3z}2 P_4}X e^{-\tfrac {3z}2 P_4},\quad
X\in\{P_i,J_{ij}\},\ i,j=1,2,3,4;
$$
explicitly,
$$
\gamma(P_i)=-P_i,\quad \gamma(J_{lm})=-J_{lm},\quad
\gamma(J_{l4})=-J_{l4}-\k_{l4}\tfrac {3z}2P_l,\quad
 i=1,2,3,4;\ \,
 l,m=1,2,3.\tag3.3
$$
\medskip

Note that the role of $\Pi_{123}$ in the Hopf homomorphisms is not
relevant: the restriction of comultiplication (3.1) to $\Pi_{123}$ gives a
non--deformed (primitive) Hopf structure. This fact is consistent with the
general structure of quantum CK algebras in which the last translation is
always taken as primitive generator [\BHOS,\BHOSii]. Thus, in general, the
essential information to deduce the quantum structure will be provided by
the CK subalgebras containing this generator and the associated $\k_i$
coefficient ($P_4$ and $\k_4$ in this case).      \medskip

\noindent
{\bf b)} {\sl\underbar {The algebra relations}}.
\medskip

The non--vanishing Lie brackets are
$$
\alignedat3
[J_{23},J_{12}]&=J_{13}, &\quad [J_{23},J_{13}]&=-\k_3J_{12} ,&\quad
[J_{12},J_{13}]&=\k_2J_{23}, \\
[J_{12},P_{1}]&=P_2 ,&\quad [J_{13},P_{1}]&=P_3, &\quad
[J_{23},P_{2}]&=P_3 ,\\
[J_{12},P_{2}]&=-\k_2P_1 ,&\quad [J_{13},P_{3}]&=-\k_2\k_3P_1 ,&\quad
[J_{23},P_{3}]&=-\k_3P_2 ,\\
[J_{12},J_{14}]&=\k_2J_{24} ,&\quad [J_{13},J_{14}]&=\k_2\k_3J_{34} ,&\quad
[J_{23},J_{24}]&=\k_3J_{34} ,\\
[J_{12},J_{24}]&=-J_{14}, &\quad [J_{13},J_{34}]&=-J_{14}, &\quad
[J_{23},J_{34}]&=-J_{24} ,\\
[J_{14},P_{4}]&=-\k_2\k_3\k_4P_1 , &\quad [J_{24},P_{4}]&=-\k_3\k_4P_2
,&\quad [J_{34},P_{4}]&=-\k_4P_3 ,
\endalignedat\tag3.4a
$$
$$
[J_{14},P_{1}]=\Sk_{-z^2}(P_4),\quad
[J_{24},P_{2}]=\Sk_{-z^2}(P_4),\quad
[J_{34},P_{3}]=\Sk_{-z^2}(P_4),\tag3.4b
$$
$$
\aligned
[J_{34},J_{24}]&=-\k_4\left\{J_{23}\C_{-z^2}(P_4)
+\tfrac{z^2}4\k_3\k_4P_1W_4\right\},\\

[J_{34},J_{14}]&=-\k_4\left\{J_{13}\C_{-z^2}(P_4)
-\tfrac{z^2}4\k_3\k_4P_2W_4\right\},\\

[J_{24},J_{14}]&=-\k_3\k_4\left\{J_{12}\C_{-z^2}(P_4)
+\tfrac{z^2}4\k_4P_3W_4\right\}.
\endaligned \tag3.4c
$$
The extra terms in (3.4c) contains $W_4$ which is a second order Casimir of
the classical subalgebra $\Pi_{123}$. The role of $q$-numbers $\q{x}={{q^x -
q^{-x}}\over{q - q^{-1}}}$ $(q=e^z)$ is played here by the so called
``generalized" trigonometric sine function [\BHOS] which, together with the
associated cosine, are given by
$$
\Sk_{-z^2}(X)=\frac{e^{z X}-e^{-z X}}{2z}, \qquad
\C_{-z^2}(X)=\frac{e^{z X}+e^{-z X}}{2}
. \tag 3.5
$$

Note that in (3.4) the six last Lie brackets are deformed ones. However,
only three of them (3.4b) remain deformed for all the 4D affine CK
algebras since their r.h.s. do not depend on the $\k_i$ parameters.
\medskip

A ``spin zero" $(W_4\equiv 0)$ differential realization of (3.4) is
given by
$$
\aligned
P_i&=\partial_{x_i},\qquad i=1,2,3,4;\\
J_{lm}&=\k_{lm} x_m\partial_{x_l}  - x_l\partial_{x_m} ,\qquad
l,m=1,2,3;\\
J_{l4}&=\k_{l4}x_4  \partial_{x_l} - x_l\Sk_{-z^2}(\partial_{x_4})
,\qquad l=1,2,3.\\
\endaligned
\tag3.6
$$

The quantum analogues of the second and fourth order Casimirs (2.4--6) are
(see Appendix A):
$$
\align
\Cal C_1^q&=4\left[\Sk_{-z^2}(\tfrac 12 P_4)\right]^2+\k_2\k_3\k_4
P_1^2+\k_3\k_4 P_2^2+\k_4P_3^2,\tag3.7\\ \Cal C_2^q&=\k_2
{W_1^q}^2+{W_2^q}^2+\k_3 {W_3^q}^2\\
&\quad +\k_3\k_4\left[\C_{-z^2}(P_4)+\tfrac{z^2}4(\k_2\k_3\k_4
P_1^2+\k_3\k_4 P_2^2+\k_4P_3^2)\right]W_4^2,\tag3.8  \endalign
$$
with
$$
\aligned
W_1^q&=-(\k_3P_2J_{34}-P_3J_{24}+\Sk_{-z^2}(P_4)J_{23}),\quad
W_2^q=\k_2\k_3P_1J_{34}-P_3J_{14}+\Sk_{-z^2}(P_4)J_{13},\\
W_3^q&=-(\k_2P_1J_{24}-P_2J_{14}+\Sk_{-z^2}(P_4)J_{12}).
\endaligned
\tag3.9
$$

Involutions and   IW contractions associated to the algebraic
structure of the classical CKG can be implemented  by
introducing a transformation of the deformation parameter $z$. So, a
$q$-involution $\Theta_{(i)}^q$ and a $q$-contraction $\Gamma_{(j)}^q$ are
defined by
$$
\align
\Theta_{(i)}^q(X,z)&:=(\Theta_{(i)}(X),-z),\ \ i=0,1,2,3;\tag3.10\\
\Gamma^q_{(j)}(X,z)&:=(\Gamma_{(j)}(X),z/\varepsilon),\ \ j=1,2,3;\tag3.11
\endalign
$$
where
$\Theta_{(i)}$ is the classical involution  (2.8) and $\Gamma_{(i)}$ the IW
contraction (2.9). The $q$-involutions (3.10) generate again an
Abelian group $(\Bbb Z_2^{\otimes 4})$ which leaves invariant the
  Hopf algebra (3.1--3.4). On the other hand, the
transformation $\Gamma^q_{(j)}$ corresponds to the limit
$\k_{j+1}\rightarrow 0$.

\bigskip \bigskip

\centerline {\bf IV. (3+1) QUANTUM KINEMATICAL AFFINE ALGEBRAS.}
\medskip\medskip

In order to identify the kinematical algebras [\BLL] as CK algebras it is
necessary to make some physical assignations to the ``geometrical"
generators $(P_i,\, J_{ij})$.
Let us take as physical generators one temporal translation $(\He)$,
three spatial translations $(\>P=(\Pe_1,\,\Pe_2,\,\Pe_3))$, three
pure inertial transformations (boosts) $(\>K=(\K_1,\,\K_2,\,\K_3))$
and three spatial rotations $(\>J=(\J_1,\,\J_2,\,\J_3))$. The assignations
we consider  are:
\medskip

\noindent
{\bf 1)} $\He=P_1$, $\>P=(P_2,P_3,P_4)$, $\>K=(J_{12},J_{13},J_{14})$,

$\>J=(J_{34},-J_{24},J_{23})$.

In this case, pure inertial transformations are non--compact only if
$\k_2\le 0$,  and Euclidean isotropy of the 3-space means $\k_3=\k_4>0$. We
therefore get the CK algebras $(0,\k_2,+,+)$, $\k_2=\{0,-\}$, which are,
respectively, Galilei  and Poincar\'e algebras. The parity ($\Cal P$) and
time--reversal  ($\Cal T$)
 automorphisms correspond, in this order, to the involutions (2.8)
$\Theta_{(1)}$ and  $\Theta_{(0)}\mdot \Theta_{(1)}$. Note that from a
speed--space contraction $(\k_2\to 0)$ of the Poincar\'e algebra we get the
Galilei algebra. \medskip

\noindent
{\bf 2)} $\He=P_4$, $\>P=(P_1,P_2,P_3)$, $\>K=(J_{14},J_{24},J_{34})$,

$\>J=(J_{23},-J_{13},J_{12})$.

Here we get $(0,+,+,\k_4)$, $\k_4=\{0,-\}$.
A Poincar\'e algebra corresponds to
$(0,+,-,-)$. By a speed--time contraction $(\k_4\to 0)$
we get a Carroll algebra $(0,+,+,0)$. The physical automorphisms are
now $\Cal P=\Theta_{(0)}\mdot \Theta_{(3)}$ and $\Cal T=\Theta_{(3)}$.
\medskip

\noindent
{\bf 3)} $\He=P_3$, $\>P=(P_4,P_1,P_2)$,
$\>K=(J_{34},J_{13},J_{23})$,  $\>J=(J_{12},J_{24},-J_{14})$.

For $(0,+,-,-)$ we obtain a Poincar\'e algebra whose
involutions are  $\Cal P=\Theta_{(0)}\mdot \Theta_{(2)}\mdot \Theta_{(3)}$
and $\Cal T=\Theta_{(2)}\mdot \Theta_{(3)}$.  \medskip

\noindent
{\bf 4)} $\He=P_2$, $\>P=(P_1,P_4,P_3)$,
$\>K=(J_{12},J_{24},J_{23})$,  $\>J=(-J_{34},-J_{13},J_{14})$.

Now, a Poincar\'e algebra is associated to $(0,-,-,+)$. Parity  and
time--reversal automorphisms are $\Theta_{(0)}\mdot \Theta_{(1)}\mdot
\Theta_{(2)}$ and $\Theta_{(1)}\mdot \Theta_{(2)}$, respectively. \medskip

The same set of assignations leads us, in the quantum case, to four
$q$-Poincar\'e and one $q$-Galilei algebras. By applying the four physical
assignations and the corresponding specializations of the parameters $\k_i$
to relations (3.1--4) we get the Hopf structure and the deformed
commutation relations. From expresions (3.7--9) we obtain the
$q$-Casimir operators. Due to its physical interest, we present the
explicit results for the Poincar\'e and Galilei quantum algebras of first
assignation.

\medskip

\noindent
{\bf 1a)} {\sl\underbar {Space--like $q$-deformed Poincar\'e algebra
$(0,-,+,+)$}}.
\medskip

\noindent
{i) Coproduct:}
$$
\aligned
\co(X)&=1\otimes X+ X\otimes 1,\qquad
X\in\{\Pe_3;\K_1,\K_2,\J_{3}\},\\
\co(\Pe_i)&=e^{-\tfrac z2 \Pe_3}\otimes \Pe_i +\Pe_i\otimes e^{\tfrac z2
\Pe_3},   \ i=1,2;\quad \co(\He)=e^{-\tfrac z2 \Pe_3}\otimes
\He +\He\otimes e^{\tfrac z2 \Pe_3},\\
\co(\K_3)&=e^{-\tfrac z2 \Pe_3}\otimes \K_3 +\K_3\otimes e^{\tfrac z2 \Pe_3}
+ \tfrac z2 \K_1e^{-\tfrac z2 \Pe_3}\otimes \Pe_1
-    \Pe_1 \otimes   e^{\tfrac z2 \Pe_3}\tfrac z2  \K_1\\
&\qquad\qquad\qquad\qquad
+ \tfrac z2 \K_2e^{-\tfrac z2 \Pe_3}\otimes \Pe_2
-    \Pe_2 \otimes   e^{\tfrac z2 \Pe_3}\tfrac z2  \K_2,\\
 \co(\J_1)&=e^{-\tfrac z2 \Pe_3}\otimes \J_1 +\J_1\otimes e^{\tfrac z2
\Pe_3} - \tfrac z2 \K_2e^{-\tfrac z2 \Pe_3}\otimes \He
+   \He \otimes   e^{\tfrac z2 \Pe_3}\tfrac z2  \K_2\\
&\qquad\qquad\qquad\qquad
- \tfrac z2 \J_3e^{-\tfrac z2 \Pe_3}\otimes \Pe_1
+     \Pe_1 \otimes   e^{\tfrac z2 \Pe_3}\tfrac z2 \J_3,\\
\co(\J_2)&=e^{-\tfrac z2 \Pe_3}\otimes \J_2 +\J_2\otimes e^{\tfrac z2 \Pe_3}
+ \tfrac z2 \K_1e^{-\tfrac z2 \Pe_3}\otimes \He
-    \He \otimes   e^{\tfrac z2 \Pe_3}\tfrac z2 \K_1\\
&\qquad\qquad\qquad\qquad
- \tfrac z2 \J_3e^{-\tfrac z2 \Pe_3}\otimes \Pe_2
+  \Pe_2 \otimes   e^{\tfrac z2 \Pe_3} \tfrac z2   \J_3.
\endaligned\tag4.1
$$

\noindent
{ii) Counit:}
$$
\epsilon(\He)=\epsilon(\Pe_i)=\epsilon(\K_{i})=\epsilon(\J_{i})=0,\quad
i=1,2,3. \tag4.2
$$

\noindent
{iii) Antipode:}
$$
\aligned
\gamma(X)&=-X,\qquad X\in\{\He,\Pe_1,\Pe_2,\Pe_3,\K_1,\K_2,\J_3\},\\
\gamma(\K_3)&=-\K_3+\tfrac {3z}2\He,\quad
\gamma(\J_1)=-\J_1-\tfrac {3z}2\Pe_2,\quad
\gamma(\J_2)=-\J_2+\tfrac {3z}2\Pe_1.
\endaligned\tag4.3
$$

\noindent
iv) Non--vanishing Lie brackets:
$$
\alignedat3
[\J_1,\J_2]&=\J_3\C_{-z^2}(\Pe_3)
+\tfrac{z^2}4 \He\, W_4,&\
[\J_2,\J_3]&= \J_1 ,&\
[\J_3,\J_1]&=\J_2 ,\\
[\J_1,\Pe_{2}]&=\Sk_{-z^2}(\Pe_3),&\
[\J_2,\Pe_{1}]&=-\Sk_{-z^2}(\Pe_3),&\
[\J_3,\Pe_{1}]&=\Pe_2 ,\\
[\J_1,\Pe_{3}]&=-\Pe_2,&\
[\J_2,\Pe_{3}]&=\Pe_1,&\
[\J_3,\Pe_{2}]&=-\Pe_1 ,\\
[\J_1,\K_2]&=\K_3,&\
[\J_2,\K_1]&=-\K_3,&\
[\J_3,\K_1]&=\K_2, \\
[\J_1,\K_3]&=- \K_2\C_{-z^2}(\Pe_3)
+\tfrac{z^2}4 \Pe_1W_4,&\ \
[\J_2,\K_3]&= \K_1\C_{-z^2}(\Pe_3)
\,+\,&\tfrac{z^2}4 \Pe_2W_4,&\ \\
[\K_i,\K_j]&=-\varepsilon_{ijk}\J_k, &\ [\K_i,\Pe_j]&=\delta_{ij}\He ,
&\ [\J_3,\K_2]&=- \K_1,\\
[\K_1,\He]&=\Pe_1,&\   [\K_2,\He]&=\Pe_2,&\ [\K_3,\He]&=\Sk_{-z^2}(\Pe_3).
\endalignedat\tag4.4
$$
where $W_4=- \He\,\J_3-\Pe_1\K_2+\Pe_2 \K_1$.
\medskip

\noindent
v) $q$-Casimirs:
$$
\Cal C_1^q=4\left[\Sk_{-z^2}(\tfrac 12 \Pe_3)\right]^2+ \Pe_1^2+
\Pe_2^2-\He^2,\tag4.5 $$
$$
\Cal C_2^q=-  {W_1^q}^2+{W_2^q}^2+
{W_3^q}^2+ \left[\C_{-z^2}(\Pe_3)+\tfrac{z^2}4(
\Pe_1^2+ \Pe_2^2-   \He^2)\right]W_4^2,\tag4.6
$$
where
$$
\aligned
W_1^q&=- \Pe_1\J_1-\Pe_2\J_2-\Sk_{-z^2}(\Pe_3)\J_3,\quad
W_2^q=-  \He\,\J_1-\Pe_2\K_3+\Sk_{-z^2}(\Pe_3)\K_2,\\
W_3^q&=-\He\,\J_2+\Pe_1\K_3-\Sk_{-z^2}(\Pe_3)\K_1.
\endaligned
\tag4.7
$$

Note that $\langle
\Pe_1,\Pe_2,\He,\K_1,\K_2,\J_{3},\Pe_3\rangle$ is a Hopf
subalgebra containing a (2+1) quantum Poincar\'e algebra as well as $\Pe_3$,
the latter being a central element. The deformation parameter $z$ has
dimensions of lenght as we can see from its behaviour under  the action of
the $q$-algebra automorphisms $q$-parity ($\Cal P^q$) and
$q$-time--reversal  ($\Cal T^q$):  $$ \aligned \Theta_{(1)}^q \equiv \Cal
P^q:&\ \{\He\rightarrow \He,\ \>P\rightarrow -\>P,\
 \>J\rightarrow \>J,\
 \>K\rightarrow -\>K;\ z\rightarrow -z  \}, \\
\Theta_{(0)}^q\mdot \Theta_{(1)}^q \equiv \Cal T^q:&\ \{\He\rightarrow
-\He,\ \>P\rightarrow \>P,\  \>J\rightarrow \>J, \  \>K\rightarrow -\>K;\
z\rightarrow z\}.
\endaligned
\tag4.8
$$
\medskip

\noindent
{\bf 1b)} {\sl\underbar {$q$-Deformed Galilei algebra $(0,0,+,+)$}}.
\medskip

It can be considered as a contraction of the above $q$-Poincar\'e algebra
when  $\k_2$ goes to zero. Its coproduct and counit are again
(4.1,2) while the antipode is
$$
\aligned
\gamma(X)&=-X,\qquad X\in\{\He,\Pe_1,\Pe_2,\Pe_3,\K_1,\K_2,\K_3,\J_3\},\\
\gamma(\J_1)&=-\J_1-\tfrac {3z}2\Pe_2,\quad
\gamma(\J_2)=-\J_2+\tfrac {3z}2\Pe_1.
\endaligned\tag4.9
$$

The main differences appear at the algebra level. Lie commutators
differing from (4.4) are
$$
[\K_i,\K_j]=0,\qquad [\K_i,\Pe_j]=0,\tag4.10
$$
and  now $W_4=-\Pe_1\K_2+\Pe_2\K_1$.

For the $q$-Casimirs we have
$$
\Cal C_1^q=4\left[\Sk_{-z^2}(\tfrac 12 \Pe_3)\right]^2+ \Pe_1^2+
\Pe_2^2,\tag4.11 $$
$$
\Cal C_2^q={W_2^q}^2+  {W_3^q}^2+ \left[\C_{-z^2}(\Pe_3)+\tfrac{z^2}4(
\Pe_1^2+ \Pe_2^2)\right]W_4^2,\tag4.12
$$
where
$$
\aligned
W_1^q&=- \Pe_1\J_1-\Pe_2\J_2-\Sk_{-z^2}(\Pe_3)\J_3,\quad

W_2^q=-\Pe_2\K_3+\Sk_{-z^2}(\Pe_3)\K_2,\\

W_3^q&=\Pe_1\K_3-\Sk_{-z^2}(\Pe_3)\K_1.
\endaligned
\tag4.13
$$

The physical $q$-automorphisms are (4.8) and $z$ has also the character of
lenght.
\medskip

\noindent
{\bf 2)} {\sl\underbar {Time--like $q$-deformed Poincar\'e algebra
$(0,+,+,-)$}}. \medskip

{}From this assignation, we exactly get the so called $\k$-Poincar\'e algebra
[\Liv]  by multiplying the basis generators by the imaginary unity $i$ and
taking   $\k=-1/z$. This deformed Poincar\'e algebra is essentially
different from the above one because the primitive generators are
$\{\He,\J_1,\J_2,\J_3\}$ and the deformation parameter has dimensions of
time. It contains as Hopf subalgebra   a quantum 3D Euclidean algebra
twisted with $\langle\He\rangle$.

\medskip

Within this scheme it is also possible to obtain a $q$-Carroll algebra
($ii'so(3)_q$) by making the contraction $\k_4\to 0$ from this
 $\k$-Poincar\'e algebra. However, the
$q$-Galilei algebra given in ref.\,[\Gill] is not a $q$-CK algebra and
strongly differs
from the studied in the first assignation (the former has a
smaller deformed algebra structure).

\bigskip

Recall that  two more $q$-Poincar\'e algebras arise in our scheme from the
third and fourth kinematical assignations. From a physical point of view
they are quite similar to the $q$-Poincar\'e analysed in {\bf 1a)} since
their  primitive generators are
$\{\Pe_1,\K_2,\K_3,\J_1\}$ and $\{\Pe_2,\K_1,\K_3,\J_2\}$
(compare with (4.1)). Moreover, in both cases the deformation
parameter $z$ can be interpreted as a lenght. It is also worth mentioning
that a 4D $q$-Euclidean $(iso(4)_q)$ algebra is got by considering the
``geometrical" generators and specializing the $q$-deformation structure to
$(0,+,+,+)$.

\medskip

Finally, it is remarkable that the method of restrictions and embeddings
used in this paper can be put forward to obtain quantum deformations for
affine CK algebras of higher dimensions. Both this work and the development
of a similar procedure for general CK algebras including semisimple ones
(all $\k_i\ne 0$) --and containing the affine CK algebras as particular
cases-- are in progress.

\bigskip
\bigskip

\centerline {\bf Acknowledgments}
\medskip

This work has been partially supported by a DGICYT project
(PB91--0196) from the Ministerio de Educaci\'on y Ciencia de
Espa\~na.

\bigskip
\bigskip

\vfil\eject

\centerline {\bf Appendix A.}
\medskip\medskip

To show that (2.5) is a fourth order Casimir it is necessary to take into
account the following non--vanishing commutation relations between the
generators and the components $W_i$ of the ``Pauli--Lubanski" vector:
$$
\alignedat4
[J_{12},W_1]&=W_2,&\quad [J_{12},W_2]&=-\k_2W_1,&\quad
[J_{13},W_3]&=-\k_2W_1,&\quad [J_{14},W_4]&=-\k_2W_1,\\
[J_{13},W_1]&=\k_3W_3,&\quad [J_{23},W_2]&=\k_3W_3,&\quad
[J_{23},W_3]&=-W_2,&\quad [J_{24},W_4]&=-W_2,\\
[J_{14},W_1]&=\k_3\k_4W_4,&\quad [J_{24},W_2]&=\k_3\k_4W_4,&\quad
[J_{34},W_3]&=\k_4W_4,&\quad [J_{34},W_4]&=-W_3.
\endalignedat\tag A.1
$$
\medskip

On the other hand, the results used to proof that $\Cal C_2^q$ is a central
element of $U_q\frak g_{(0,\k_2,\k_3,\k_4)}$, which include the
non--vanishing  commutation relations between the components $(W_i^q,W_4)$
of $\Cal C_2^q$ and the basis generators,  are the following:
$$
\align
[J_{12},W_1^q]&=W_2^q,\qquad\quad
[J_{24},W_1^q]=-\tfrac {z^2}4\k_3^2\k_4^2P_1P_2W_4,\\
[J_{13},W_1^q]&=\k_3 W_3^q,\qquad
[J_{34},W_1^q]=-\tfrac {z^2}4 \k_3\k_4^2P_1P_3W_4,\tag A.2\\
[J_{14},W_1^q]&=\k_3\k_4
\left[\C_{-z^2}(P_4)+\tfrac{z^2}4(\k_3\k_4 P_2^2+\k_4P_3^2)\right]W_4,
\endalign
$$
$$
\align
[J_{12},W_2^q]&=-\k_2W_1^q,\quad\,
[J_{14},W_2^q]=-\tfrac {z^2}4\k_2\k_3^2\k_4^2P_1P_2W_4,\\
[J_{23},W_2^q]&=\k_3 W_3^q,\qquad
[J_{34},W_2^q]=-\tfrac {z^2}4 \k_3\k_4^2P_2P_3W_4,\tag A.3\\
[J_{24},W_2^q]&=\k_3\k_4
\left[\C_{-z^2}(P_4)+\tfrac{z^2}4(\k_2\k_3\k_4 P_1^2+\k_4P_3^2)\right]W_4,
\endalign
$$
$$
\align
[J_{13},W_3^q]&=-\k_2W_1^q,\quad
[J_{14},W_3^q]=-\tfrac {z^2}4\k_2\k_3\k_4^2P_1P_3W_4,\\
[J_{23},W_3^q]&=- W_2^q,\qquad
[J_{24},W_3^q]=-\tfrac {z^2}4 \k_3\k_4^2P_2P_3W_4,\tag A.4\\
[J_{34},W_3^q]&=\k_4
\left[\C_{-z^2}(P_4)+\tfrac{z^2}4(\k_2\k_3\k_4
P_1^2+\k_3\k_4P_2^2)\right]W_4, \endalign $$
$$
\align
[J_{14},W_4]&=-\k_2W_1^q,\quad [W_4,W_1^q]=-\k_3P_2W_3^q+P_3W_2^q,\\
 [J_{24},W_4]&=-W_2^q,\qquad
[W_4,W_2^q]=\k_2\k_3P_1W_3^q-\k_2P_3W_1^q,\tag A.5\\
[J_{34},W_4]&=-W_3^q,\qquad [W_4,W_3^q]=-\k_2P_1W_2^q+\k_2P_2W_1^q,
\endalign
$$
$$
[\Sk_{-z^2}(P_4),J_{i4}]=\k_{i4}P_i\C_{-z^2}(P_4),\quad
[\C_{-z^2}(P_4),J_{i4}]=z^2\k_{i4}P_i\Sk_{-z^2}(P_4),\quad i=1,2,3.
\tag A.6
$$
\bigskip
\bigskip

\vfil\eject

\centerline {\bf Appendix B.}
\medskip\medskip

The coproduct and the deformed Lie brackets corresponding
to the  3D affine $q$-CK algebras $(U_q\frak g_{(0,\k_2,\k_3)}=\langle
P_1,P_2,P_3,J_{12},J_{13},J_{23}\rangle)$, using the results of
ref. [\BHOSii], are:
$$
\aligned
\co(P_3)&=1\otimes P_3 + P_3\otimes 1,\qquad\qquad\quad
\co(J_{12})=1\otimes J_{12} + J_{12}\otimes 1,\\
\co(P_1)&=e^{-{z\over 2}P_3} \otimes
P_1 + P_1 \otimes e^{{z\over2}P_3} ,\quad
\co(P_2)=e^{-{z\over 2}P_3}\otimes
P_2 + P_2 \otimes e^{{z\over2}P_3} ,\\
\co(J_{13})&=e^{-{z\over 2}P_3}\otimes
J_{13} + J_{13} \otimes e^{{z\over2}P_3} + \tfrac z 2 J_{12}e^{-{z\over
2}P_3} \otimes \k_3 P_2 - \k_3 P_2
\otimes e^{{z\over2}P_3}\tfrac z 2 J_{12},\\
\co(J_{23})&=e^{-{z\over 2}P_3} \otimes
J_{23} + J_{23} \otimes e^{{z\over2}P_3}  -\tfrac z 2 J_{12} e^{-{z\over
2}P_3} \otimes \k_3 P_1 + \k_3
P_1\otimes e^{{z\over2}P_3}\tfrac z 2 J_{12}.
\endaligned
\tag B.1
$$
$$
[J_{23},J_{13}]=-\k_3 J_{12}\C_{-z^2}(P_3),\quad
[J_{23},P_2]=\Sk_{-z^2}(P_3),\quad
[J_{13},P_1]=\Sk_{-z^2}(P_3),\tag B.2
$$
\medskip

 We write down this structure for the  subalgebras $\Pi_{124}$, $\Pi_{134}$
and $\Pi_{234}$ (2.10):
\medskip

\noindent
(a) $\Pi_{124}\leftrightarrow\langle
P_1,P_2,P_4,J_{12},J_{14},J_{24}\rangle,\  (0,\k_2,\k_3\k_4)$.
$$
\aligned
\co(P_4)&=1\otimes P_4 + P_4\otimes 1,\qquad\qquad\quad
\co(J_{12})=1\otimes J_{12} + J_{12}\otimes 1,\\
\co(P_1)&=e^{-{z\over 2}P_4} \otimes
P_1 + P_1 \otimes e^{{z\over2}P_4} ,\quad
\co(P_2)=e^{-{z\over 2}P_4}\otimes
P_2 + P_2 \otimes e^{{z\over2}P_4} ,\\
\co(J_{14})&=e^{-{z\over 2}P_4}\otimes
J_{14} + J_{14} \otimes e^{{z\over2}P_4} + \tfrac z 2 J_{12}e^{-{z\over
2}P_4} \otimes \k_3\k_4 P_2 - \k_3\k_4 P_2
\otimes e^{{z\over2}P_4}\tfrac z 2 J_{12},\\
\co(J_{24})&=e^{-{z\over 2}P_4} \otimes
J_{24} + J_{24} \otimes e^{{z\over2}P_4}  -\tfrac z 2 J_{12} e^{-{z\over
2}P_4} \otimes \k_3\k_4 P_1 + \k_3\k_4
P_1\otimes e^{{z\over2}P_4}\tfrac z 2 J_{12}.
\endaligned
\tag B.3
$$
$$
[J_{24},J_{14}]=-\k_3\k_4 J_{12}\C_{-z^2}(P_4),\quad
[J_{24},P_2]=\Sk_{-z^2}(P_4),\quad
[J_{14},P_1]=\Sk_{-z^2}(P_4),\tag B.4
$$
\medskip

\noindent
(b) $\Pi_{134}\leftrightarrow\langle
P_1,P_3,P_4,J_{13},J_{14},J_{34}\rangle,\  (0,\k_2\k_3,\k_4)$.
$$
\aligned
\co(P_4)&=1\otimes P_4 + P_4\otimes 1,\qquad\qquad\quad
\co(J_{13})=1\otimes J_{13} + J_{13}\otimes 1,\\
\co(P_1)&=e^{-{z\over 2}P_4} \otimes
P_1 + P_1 \otimes e^{{z\over2}P_4} ,\quad
\co(P_3)=e^{-{z\over 2}P_4}\otimes
P_3 + P_3 \otimes e^{{z\over2}P_4} ,\\
\co(J_{14})&=e^{-{z\over 2}P_4}\otimes
J_{14} + J_{14} \otimes e^{{z\over2}P_4} + \tfrac z 2 J_{13}e^{-{z\over
2}P_4} \otimes \k_4 P_3 - \k_4 P_3
\otimes e^{{z\over2}P_4}\tfrac z 2 J_{13},\\
\co(J_{34})&=e^{-{z\over 2}P_4} \otimes
J_{34} + J_{34} \otimes e^{{z\over2}P_4}  -\tfrac z 2 J_{13} e^{-{z\over
2}P_4} \otimes \k_4 P_1 + \k_4
P_1\otimes e^{{z\over2}P_4}\tfrac z 2 J_{13}.
\endaligned
\tag B.5
$$
$$
[J_{34},J_{14}]=-\k_4 J_{13}\C_{-z^2}(P_4),\quad
[J_{34},P_3]=\Sk_{-z^2}(P_4),\quad
[J_{14},P_1]=\Sk_{-z^2}(P_4),\tag B.6
$$
\medskip

\noindent
(c) $\Pi_{234}\leftrightarrow\langle
P_2,P_3,P_4,J_{23},J_{24},J_{34}\rangle,\  (0,\k_3,\k_4)$.
$$
\aligned
\co(P_4)&=1\otimes P_4 + P_4\otimes 1,\qquad\qquad\quad
\co(J_{23})=1\otimes J_{23} + J_{23}\otimes 1,\\
\co(P_2)&=e^{-{z\over 2}P_4} \otimes
P_2 + P_2 \otimes e^{{z\over2}P_4} ,\quad
\co(P_3)=e^{-{z\over 2}P_4}\otimes
P_3 + P_3 \otimes e^{{z\over2}P_4} ,\\
\co(J_{24})&=e^{-{z\over 2}P_4}\otimes
J_{24} + J_{24} \otimes e^{{z\over2}P_4} + \tfrac z 2 J_{23}e^{-{z\over
2}P_4} \otimes \k_4 P_3 - \k_4 P_3
\otimes e^{{z\over2}P_4}\tfrac z 2 J_{23},\\
\co(J_{34})&=e^{-{z\over 2}P_4} \otimes
J_{34} + J_{34} \otimes e^{{z\over2}P_4}  -\tfrac z 2 J_{23} e^{-{z\over
2}P_4} \otimes \k_4 P_2 + \k_4
P_2\otimes e^{{z\over2}P_4}\tfrac z 2 J_{23}.
\endaligned
\tag B.7
$$
$$
[J_{34},J_{24}]=-\k_4 J_{23}\C_{-z^2}(P_4),\quad
[J_{34},P_3]=\Sk_{-z^2}(P_4),\quad
[J_{24},P_2]=\Sk_{-z^2}(P_4).\tag B.8
$$

\bigskip\bigskip

\vfil\eject

\centerline {\bf References}

\ninepoint\medskip

\ref      %% 1
\no[{\Dr}]
\by V.\, G.\, Drinfeld
\paper Quantum Groups
\jour Proceedings of the International Congress of Mathematics,
MRSI Berkeley, (1986) 798
\endref

\ref      %% 2
\no[{\Ji}]
\by M. Jimbo
\jour Lett. Math. Phys.
\yr 1985
\vol 10
\pages 63

\moreref
\paper \;
\yr 1986
\vol 11
\pages 247
\endref

\ref     %% 3
\no[{\Li}]
\by       J. Lukierski, H. Ruegg, A. Nowicky and V.N. Tolstoi
\jour     Phys. Lett. B.
\vol      264
\yr       1991
\pages    331
\endref

\ref     %% 4
\no[{\Lii}]
\by       J. Lukierski, H. Ruegg and A. Nowicky
\jour     Phys. Lett. B.
\vol      271
\yr       1991
\pages    321
\endref

\ref     %% 5
\no[{\Liii}]
\by       J. Lukierski and A. Nowicky
\jour     Phys. Lett. B.
\vol      279
\yr       1992
\pages    299
\endref

\ref     %% 6
\no[{\Liv}]
\by       J. Lukierski, A. Nowicky and H. Ruegg
\jour     Phys. Lett. B.
\vol      293
\yr       1992
\pages    344
\endref

\ref     %% 7
\no[{\Ti}]
\by      E. Celeghini, R. Giachetti, E. Sorace and M. Tarlini
\paper   Contractions of quantum groups
\jour    Lecture Notes in Mathematics n. 1510, 221,
         Springer-Verlag, Berl\1n 1992
\endref

\ref        %%  8
\no[{\BHOS}]
\by         A. Ballesteros, F.J. Herranz, M.A. del Olmo and M. Santander
\paper      Quantum structure of the motion groups of the two--dimensional
            Cayley--Klein geometries
\jour       J. Phys. A, to appear
\vol
\yr
\endref

\ref        %%  9
\no[{\BHOSii}]
\by         A. Ballesteros, F.J. Herranz, M.A. del Olmo and M. Santander
\paper      Quantum (2+1) kinematical algebras: a global approach
\jour       UVA, preprint, Valladolid
\vol
\yr         1993
\endref

\ref       %% 10
\no[{\IW}]
\by        E. In\"on\"u, E. P. Wigner
\jour      Proc. Natl. Acad. Sci. U. S.
\vol       39
\yr        1953
\pages     510
\endref

\ref      %% 11
\no[{\BLL}]
\by        H. Bacry and J. M. L\'evy--Leblond
\jour      J. Math. Phys.
\vol       9
\yr        1968
\pages     1605
\endref

\ref     %% 12
\no[{\Gill}]
\by       S. Giller, P. Koshinski, M. Majewski, P. Maslanka and J. Kunz
\jour     Phys. Lett. B.
\vol      286
\yr       1992
\pages    57
\endref

\ref      %% 13
\no[{\TesHe}]
\book     Geometr\1as de Cayley--Klein en N dimensiones y grupos
          cinem\'aticos
\by       F.J. Herranz
\yr       1991
\bookinfo Tesina de Licenciatura, Universidad de Valladolid
\endref

\ref      %% 14
\no[{\SHO}]
\by       M. Santander, F.J. Herranz and M. A. del Olmo
\paper    Kinematics and homogeneous spaces for symmetrical contractions
          of orthogonal groups
\jour     Proceedings of the  XIX Int. Colloq. on Group Theor. Methods in
          Phys.,  CIEMAT, Madrid, (1993)
\endref

\ref     %% 15
\no[{\HOS}]
\by        F.J. Herranz, M. A. del Olmo and M. Santander
\jour      in preparation
%\vol
%\yr
\pages
\endref

\ref      %% 16
\no[{\HMOS}]
\book     Cayley--Klein geometries and graded contractions of $so(N+1)$
\by       F.J. Herranz, M. de Montigny, M.A. del Olmo and M. Santander
\yr       (1993)
\bookinfo UVA/CRM preprint
\endref

\ref     %% 17
\no[{\Abe}]
\by         E. Abe
\book       Hopf Algebras
\publ       Cambridge Tracts in Mathematics 74, Cambridge University. Press
Cambridge 1980
\publaddr
\yr
\endref

\ref     %% 18
\no[{\Maj}]
\by      S. Majid
\jour    Int. Jour. Mod. Phys. A
\vol     5
\yr      1991
\pages   1
\endref

\end